# Making LLMs Work for Enterprise Data Tasks

Çağatay Demiralp[1]  Fabian Wenz[1]  Peter Baile Chen[1]  Moe Kayali[2]  Nesime Tatbul[1,3]  Michael Stonebraker[1]
[1]MIT  [2]UW  [3]Intel

Large language models (LLMs) have shown strong performances on natural language (NL) comprehension tasks, from summarization to question answering. The power of these models comes from optimizing for simple self-supervised learning tasks such as next token prediction using massive public web texts as training data on a scalable and adaptive architecture. However, by construction, LLMs know little about enterprise database tables in the private data ecosystem, which differ substantially from web text in structure and content.

**Contributions.** Given LLMs' performance is tied to their training data [1], a crucial question is how useful they can be in improving enterprise database management and analysis tasks. To help contend with this question, we contribute (1) preliminary experimental results on the performance of LLMs for *text-to-SQL* and *semantic column-type detection* tasks on enterprise datasets and (2) a discussion of challenges and potential solutions for effectively utilizing LLMs in enterprise settings.

**Experiment I**

**Task.** In our first experiment, we consider the text-to-SQL task. This task aims to generate the corresponding SQL query for a given question in natural language. It has garnered considerable attention due to its potential to significantly improve the usability and accessibility of data management and analysis systems.

**Dataset.** We create a benchmark containing 37 NL questions—SQL query pairs on 99 tables from the MIT Data Warehouse (DW). The MIT DW is the central data store that combines data about MIT faculty, facilities, and students. Our benchmark is an enterprise analog of text-to-SQL benchmarks such as Spider [2] that use public data.

**Setup & Metrics**. To generate the SQL query corresponding to a given NL question, we use GPT 4.0 through prompting (in-context learning). We provide no examples (zero-shot) but include different types of context: *schema* (the schema definitions of all the tables in the dataset), *querylog* (prior queries run on each table), and *rules* (rules and facts specific to the dataset and its domain). We evaluate the performance of the generated query with the accuracy of the table retrieval step. We consider each query generated by the LLM a single prediction. A prediction is correct if the tables retrieved are identical to those used in the ground-truth SQL query.

**Results.** We obtain the accuracies of 0% with *no context* (baseline sanity check), 40% with *schema*, 40% with *schema+rules,* and 48% with *schema+querylog*.

**Discussion.** These performance numbers are significantly lower than typical text-to-SQL performances of LLMs on public data benchmarks such as Spider [2]. For example, we get an accuracy of 42% with *no context* provided for 37 NL questions sampled from Spider. GPT 4.0 must have seen the benchmark during pretraining. Providing prior queries in context considerably increases the accuracy, suggesting a promising direction. We are currently extending our benchmark dataset with new questions, including those provided by DW admins and tables from new enterprise DWs. We also collect rules from DW admins to consider in our ongoing experiments.

**Experiment II**

**Task.** In our second experiment, we evaluate LLMs' semantic column type detection (or annotation) performance on enterprise data. The goal of this task is to assign a semantic label, a real-world reference ("population," "address," "salary," etc.), to a given table column, characterizing the values in the column.

**Dataset.** We use an enterprise dataset (Goby) with 1,187 tables containing information about different forms of events, such as concerts and shows. Each table contains information about events from a particular domain (arts, entertainment, etc.), and the table schemas are heterogeneous. In addition, human labelers have manually inspected these tables to assign a semantic type to each column out of a domain-specific ontology. Finally, the labelers have also constructed a universal schema that unifies data from all source tables.

**Setup & Metrics.** We used GPT 3.5-turbo through prompting to predict semantic types without providing examples (zero-shot). We ask the LLM to predict the class of a given serialized table column by picking one of the 34 ground-truth types. We measure the performance on each target semantic type by calculating the F1, precision, and recall scores. We also measure the ability of the LLM to suggest semantic labels by inspecting the data and comparing the suggested (discovered) labels to the ground truth labels.

**Results.** Using a zero-shot approach, we obtained a class-weighted F1 score of 71%, precision of 77%, and recall of 70% across the 1,187. Separately, when generating semantic labels *ab initio* from the data values with an LLM, we find that the LLM can recover 70.5% of the labels the human raters devised. Finally, the LLM can create a plausible property hierarchy (ontology) encompassing all the Goby dataset's semantic types.

**Discussion.** The F1, precision, and recall on this private dataset are about 10% lower than those reported with LLM-based approaches on public data [3]. The results suggest a significant data distribution shift between public and enterprise data. The good performance on ontology generation shows promise for specializing semantic labels to private datasets.

### Challenges

Our initial findings and conversations with industry practitioners have highlighted several limitations of LLMs that will impede their utility for enterprise data tasks:

**Latency**. Utilizing LLMs via API calls at scale introduces latency issues, undermining the performance and scalability required by enterprise applications. The brittle latency patterns are a nonstarter for many interactive use cases (e.g., conversational agents).

**Cost**. LLMs' computational demands, particularly when leveraging GPUs for training and inference and using them through model-as-a-service APIs, present a significant financial burden for many enterprises. The performance of LLMs may not justify their costs, as there are cheaper, smaller, and faster models that can perform the same task.

**Quality**. Challenges such as evaluation (e.g., automated evaluation at scale), non-determinism, lack of reproducibility, explainability, and control undermine the reliability and trustworthiness of LLM-based approaches. Unsurprisingly, LLM hallucinations are also a big problem with enterprise data. The low accuracy scores we have seen so far are not conducive to adopting LLMs in enterprise settings.

### Possible Solutions

**A New Toolstack**. We need a new set of tools that effectively combine the high recall properties of LLMs with the high precision of rules (e.g., code) and cost and latency-effective local models, going beyond current orchestration and retrieval augmented generation (RAG) frameworks.

**Representation Learning Models for Enterprise Data Systems**. The current practice of nudging the behavior of an LLM as a black box through prompting to achieve desired results is inherently ineffective. We must develop representation learning models specific to enterprise data systems with relevant pretraining tasks. Enterprise data residing in data systems not only has different semantic, statistical, and structural characteristics but also has a much richer context [4]. We know how data is used or acted upon (think of query, application, and interaction logs) and by whom. Enabling models to train on both data and "action" context will improve the precision of learned representations. However, a challenge in this direction is the availability of database tables at scale for training; we believe data synthesis using simulation needs to be part of the solution.

### References

[1] Kandpal et al. Large Language Models Struggle to Learn Long-Tail Knowledge. *ICML*, 2023.
[2] Yu et al. Spider: A Large-Scale Human-Labeled Dataset for Complex and Cross-Domain Semantic Parsing and Text-to-SQL Task. *EMNLP*, 2018.
[3] Kayali et al. CHORUS: Foundation Models for Unified Data Discovery and Exploration. *VLDB*, 2024.
[4] Hellerstein et al. Ground: A Data Context Service. *CIDR*, 2017.